\pdfoutput=1
\documentclass[manuscript]{acmart}
\AtBeginDocument{%
  \providecommand\BibTeX{{%
    \normalfont B\kern-0.5em{\scshape i\kern-0.25em b}\kern-0.8em\TeX}}}

\copyrightyear{2022}
\acmYear{2022}
\setcopyright{rightsretained}
\acmConference[FAccT '22]{2022 ACM Conference on Fairness, Accountability, and Transparency}{June 21--24, 2022}{Seoul, Republic of Korea}
\acmBooktitle{2022 ACM Conference on Fairness, Accountability, and Transparency (FAccT '22), June 21--24, 2022, Seoul, Republic of Korea}\acmDOI{10.1145/3531146.3533229} \acmISBN{978-1-4503-9352-2/22/06}

\usepackage[english]{babel}

\usepackage{amsmath}
\usepackage{graphicx}
\usepackage{spverbatim}
\usepackage{fancyvrb}
\usepackage{csquotes}
\usepackage[T1]{fontenc}
\usepackage[utf8]{inputenc}
\hypersetup{linkcolor=blue,filecolor=magenta,urlcolor=cyan} 
\usepackage[labelfont=bf]{caption}
\usepackage[font={small}]{caption}
\title{Predictability and Surprise in Large Generative Models 
}

\begin{document}

\author{Deep Ganguli}
\email{deep@anthropic.com}
\authornote{Core Research Contributors}
\author{Danny Hernandez}
\email{danny@anthropic.com}
\authornotemark[1]
\author{Liane Lovitt}
\email{liane@anthropic.com}
\authornotemark[1]
\author{Nova DasSarma}
\email{nova@anthropic.com}
\authornote{Core Infrastructure Contributors}
\author{Tom Henighan}
\email{henighan@anthropic.com}
\authornotemark[2]
\author{Andy Jones}
\email{andy@anthropic.com}
\authornotemark[2]
\author{Nicholas Joseph}
\email{nick@anthropic.com}
\authornotemark[2]
\author{Jackson Kernion}
\email{jackson@anthropic.com}
\authornotemark[2]
\author{Ben Mann}
\email{ben@anthropic.com}
\authornotemark[2]
\author{Amanda Askell}
\email{amanda@anthropic.com}
\author{Yuntao Bai}
\email{yuntao@anthropic.com}
\author{Anna Chen}
\email{anna@anthropic.com}
\author{Tom Conerly}
\email{tom.conerly@anthropic.com}
\author{Dawn Drain}
\email{dawn@anthropic.com}
\author{Nelson Elhage}
\email{nelhage@anthropic.com}
\author{Sheer El Showk}
\email{sheer@anthropic.com}
\author{Stanislav Fort}
\email{stan@anthropic.com}
\author{Zac Hatfield-Dodds}
\email{zac@anthropic.com}
\author{Scott Johnston}
\email{scott@anthropic.com}
\author{Shauna Kravec}
\email{shauna@anthropic.com}
\author{Neel Nanda}
\email{neel@anthropic.com}
\author{Kamal Ndousse}
\email{kamal@anthropic.com}
\author{Catherine Olsson}
\email{catherio@anthropic.com}
\author{Daniela Amodei}
\email{daniela@anthropic.com}
\author{Tom Brown}
\email{tom@anthropic.com}
\author{Jared Kaplan}
\email{jared@anthropic.com}
\author{Sam McCandlish}
\email{sam@anthropic.com}
\author{Chris Olah}
\email{colah@anthropic.com}
\author{Dario Amodei}
\email{dario@anthropic.com}
\author{Jack Clark}
\authornote{Correspondence to: jack@anthropic.com} 
\authornote{Author contributions are listed in Appendix \ref{app:author_contributions}}
\email{jack@anthropic.com}
\affiliation{
\institution{Anthropic}
\city{San Francisco}
\country{USA}
}

\renewcommand{\shortauthors}{Ganguli, et al.}

\begin{abstract}
Large-scale pre-training has recently emerged as a technique for creating capable, general-purpose, generative models such as GPT-3, Megatron-Turing NLG, Gopher, and many others. In this paper, we highlight a counterintuitive property of such models and discuss the policy implications of this property. Namely, these generative models have a paradoxical combination of predictable loss on a broad training distribution (as embodied in their "scaling laws"), and unpredictable specific capabilities, inputs, and outputs. We believe that the high-level predictability and appearance of useful capabilities drives rapid development of such models, while the unpredictable qualities make it difficult to anticipate the consequences of model deployment. We go through examples of how this combination can lead to socially harmful behavior with examples from the literature and real world observations, and we also perform two novel experiments to illustrate our point about harms from unpredictability. Furthermore, we analyze how these conflicting properties combine to give model developers various motivations for deploying these models, and challenges that can hinder deployment. We conclude with a list of possible interventions the AI community may take to increase the chance of these models having a beneficial impact. We intend this paper to be useful to policymakers who want to understand and regulate AI systems, technologists who care about the potential policy impact of their work, funders who want to support work addressing these challenges, and academics who want to analyze, critique, and potentially develop large generative models.
\end{abstract}

\maketitle

\section{Introduction}

 Scaling up the amount of data, compute power, and model parameters of neural networks has recently led to the arrival (and real world deployment) of capable generative models such as CLIP \cite{radford_learning_2021}, Ernie $3.0$ Titan \cite{wang_ernie_2021}, FLAN \cite{wei_finetuned_2021}, Gopher \cite{rae_scaling_2021}, GPT-3 \cite{brown_language_2020}, HyperClova \cite{kim_what_2021}, Jurassic-1-Jumbo \cite{lieber_jurassic-1_2021}, Megatron Turing NLG \cite{smith_using_2022}, LaMDA \cite{thoppilan_lamda_2022}, Pan Gu \cite{zeng_pangu-alpha_2021}, Yuan $1.0$ \cite{wu_yuan_2021}, and more. For this class of models\footnote{Some refer to this class of models as \enquote*{foundation models} \cite{bommasani_opportunities_2021}.} the relationship between scale and model performance is often so predictable that it can be described in a lawful relationship --- a scaling law. In most cases, these scaling laws predict a continued increase in certain capabilities as models get larger. At the same time, larger generative models represent an increasing proportion of the eye-catching results in machine learning. As a result, many institutions have started producing large models over the past few years, in response to the predictability afforded by scaling laws, and the fact these models can be plugged into systems that generate economic value, like search engines.\footnote{We do not discuss to whom this economic value accrues, and we do not intend to imply that by default it will accrue broadly or that no one will be harmed.} It has also  become clear that these models present novel risks of harmful behavior, which are difficult to predict and may become more severe as the models increase in capability. Attempts to study these harms with smaller models may not accurately reflect what occurs in larger ones.

In this paper, we attempt to better understand the influence of scaling laws on the dynamics of large-scale model development and deployment, with a focus on large language models. \textbf{Our basic thesis is that large generative models have a paradoxical combination of high predictability --- model loss improves in relation to resources expended on training, and tends to correlate loosely with improved performance on many tasks --- and high unpredictability --- specific model capabilities, inputs, and outputs can’t be predicted ahead of time. The former drives rapid development of such models while the latter makes it difficult to anticipate the consequences of their development and deployment.}  We go through examples of how this combination can lead to socially harmful behavior, while also analyzing the motivations and challenges that developers of such models will face. Our goal in this paper is to outline how and why we expect these models to be developed, so we can identify interventions to guide model development.  We conclude with some policy recommendations that could increase the safety of large-scale model deployments, and improve the incentive structure for developers building these models. Though all of the individual points about scaling laws, open-endedness, or the proliferation of large models are explicitly or implicitly presented in other research, our contribution here is to highlight the complete picture together with its implications.

Although we focus on scaling laws, many of our points complement related work on the societal risks of deploying large models \cite{bender_dangers_2021, tamkin_understanding_2021, bommasani_opportunities_2021, dinan_anticipating_2021, weidinger_ethical_2021, kenton_alignment_2021}. However, similarly to \cite{weidinger_ethical_2021}, we do not consider here the costs of human labor involved in creating and annotating training data \cite{gray_ghost_2019}, the ethics of supply chains involved in creating the requisite hardware on which to train models \cite{crawford_atlas_2021}, or the environmental costs of training models \cite{bender_dangers_2021, patterson_carbon_2021, schwartz_green_2020, strubell_energy_2019}. Scaling laws are likely to significantly impact these issues.

The remainder of the paper is organized as follows. In Section \ref{sec:1}, we articulate and support our central thesis about large generative models by decomposing it into four claims, each of which we support with evidence from previously published data, and in some cases, with novel experiments on large language models \cite{askell_general_2021}. In Section \ref{sec:1.1} we discuss smooth general capability scaling. More precisely, by general capability scaling we mean two things. First, the training (and test) loss improves predictably with scale on a broad data distribution.  Second, this improvement in loss tends to correlate on average with increased performance on a number of downstream tasks \cite{brown_language_2020, rae_scaling_2021}. We refer to the combination of these two properties throughout the paper as smooth general capability (or performance) scaling.\footnote{Note that, as will be discussed later as the central thesis of the paper, smooth general capability scaling does not imply smooth scaling on any particular task. It also does not imply that the tasks typically measured are the only tasks that are important; indeed the presence of unmeasured tasks is part of our thesis.}
 In Section \ref{sec:1.2}, we discuss abrupt specific capability scaling, in which models can also suddenly gain specific capabilities at scale. We illustrate this phenomenon with three examples from the literature \cite{brown_language_2020, rae_scaling_2021, austin_program_2021}. In Section \ref{sec:1.3}, we argue that entire areas of model competency may be unknown until they are solicited from specific inputs, problem domains, or applications. In Section \ref{sec:1.4}, we discuss challenges that arise from the open-endedness of model outputs and show both qualitative and quantitative examples of harmful and toxic outputs emerging with scale.

\begin{figure}[t]
\centering
\includegraphics[width=0.99\textwidth]{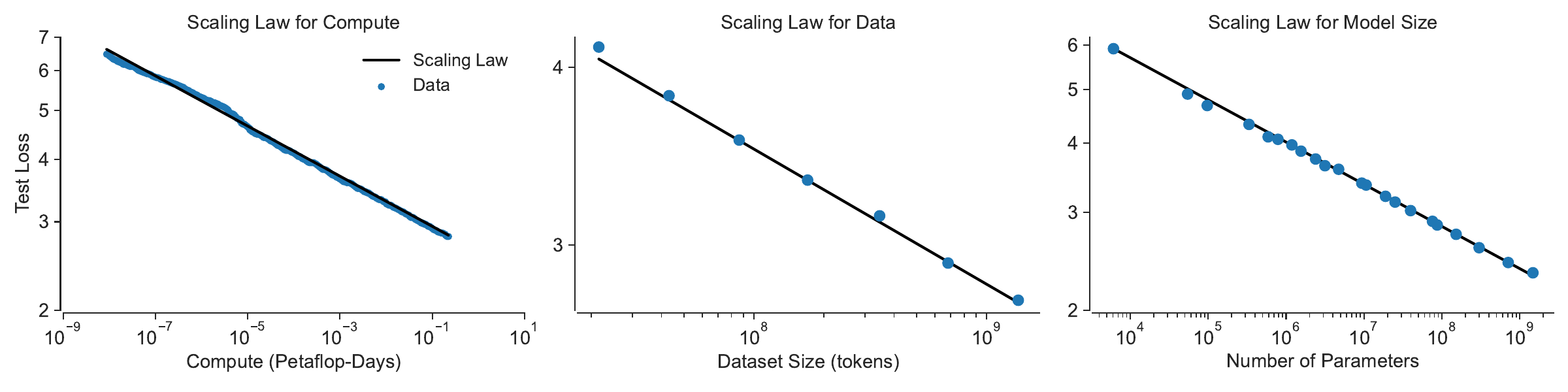}
\vspace{-.5em}
\caption{\label{fig:scaling_laws} Scaling laws reliably predict that model performance (y-axes) improves with increasing compute \textbf{(Left)}, training data \textbf{(Middle)}, and model size \textbf{(Right)}. In all cases a power-law (straight line, black) fits the empirically observed data (blue) exceptionally well. Figure adapted from \cite{kaplan_scaling_2020}.}
\vspace{-1em}
\end{figure}

In Section \ref{sec:2}, we outline why, despite these conflicting properties of predictability and unpredictability, we expect increasing development and deployment of large generative models despite the challenges we outline in Section \ref{sec:1}. We posit that this is due to a confluence of economic, scientific, and prestige motivations, each of which we summarize. We also consider a few possible barriers to entry that model developers may face during development and deployment, including high financial costs, access to engineering talent, safety concerns, and a lack of standards on how to responsibly deploy capable generative models. We also provide some empirical observations (grounded in the motivations and challenges described above) about how the development of large language models has unfolded thus far, including a quantitative analysis of the increasing gap between academia and industry for large model development.

Finally, in Section \ref{sec:3} we outline policy interventions that may help concretely address the challenges we outline in Sections \ref{sec:1} and \ref{sec:2} in order to help guide the development and deployment of larger models for the broader social good. We leave some illustrative experiments, technical details, and caveats about our claims in Appendix \ref{app}.

\section{Distinguishing Features of Large Generative Models} \label{sec:1}

We claim that large generative models (e.g., GPT-3 \cite{brown_language_2020}, LaMDA \cite{thoppilan_lamda_2022}, Gopher \cite{rae_scaling_2021}, etc.) are distinguished by four features:

\begin{itemize}
    \item 
    \textbf{Smooth, general capability scaling}: It is possible to \emph{predictably} improve the {general} performance of generative models --- their loss on capturing a specific, though very broad, data distribution --- by scaling up the size of the models, the compute used to train them, and the amount of data they’re trained on in the correct proportions. These proportions can be accurately predicted by scaling laws (Figure \ref{fig:scaling_laws}). We believe that these scaling laws de-risk investments in building larger and generally more capable models despite the high resource costs and the difficulty of predicting precisely how well a model will perform on a specific task. Note, the harmful properties of models, such as toxicity, can scale alongside directly helpful capabilities. 
    
    \item
    \textbf{Abrupt, specific capability scaling}: Though performance is predictable at a general level, performance on a specific task can sometimes emerge quite unpredictably and abruptly at scale.\footnote{Similar behavior has also been observed during the training process of an individual model (rather than as a function of model size) for algorithmic tasks, and has been termed “grokking” \cite{power_grokking_2022}.}  While counter-intuitive, this is possible because any specific task is a tiny slice of a model's output probability distribution, and so can change rapidly even as the full distribution remains smooth. 
    
    \item
    \textbf{Open-ended inputs and domains}: Large generative models are open-ended and can take in a varying range of inputs concerning arbitrary domains. As a result, certain capabilities (or even entire areas of competency) may be unknown until an input happens to be provided that solicits such knowledge. Even after a model is trained, creators and users may not be aware of most of its (possibly harmful) capabilities. These properties become more pronounced as the models scale ---  larger models tend to be harder to characterize than smaller ones. 
    
    \item
    \textbf{Open-ended outputs}: Finally, model outputs are also open-ended in the sense that they are difficult to predict or control, even given a fixed scale, input, topic, or task. These outputs may be helpful or harmful, but it's difficult to know in advance. Of course, models with both open-ended inputs and outputs have existed for decades, but what is new is the level of capability and breadth of open-endedness.

\end{itemize}

 \noindent{In the following sections, we further describe each of these distinguishing features, and discuss how combinations of them may lead to disruptive societal impacts. We support our claims with data and experiments.}

\subsection{Smooth General Capability Scaling} \label{sec:1.1}
Generally, machine learning experiments are not precisely predictable --- complex models trained on complex data typically yield noisy or variable results \cite{zhuang_randomness_2021, clary_lets_2019}.\footnote{For example, \cite{clary_lets_2019} documents a strong lack of run-to-run reproducibility in reinforcement learning on Atari games when only changing the initial random seed. This suggests that differences between algorithms may be difficult to measure rigorously due to such intrinsic noise.} Though individual experiments may be unpredictable, the general performance of large generative models tends to exhibit smooth and predictable growth as a function of scale --- larger systems tend to do increasingly better on a broad range of tasks. This was first noticed by  \cite{hestness_deep_2017} who observed that capabilities such as machine translation and speech recognition increased in a smooth, predictable manner as the size of the model increased. Subsequent work formalized and experimentally validated a quantitative relationship between scale (in terms of both model size and training data size) and model generalization error \cite{rosenfeld_constructive_2019}. Furthermore,  \cite{kaplan_scaling_2020} demonstrated that test loss performance on language modeling tasks scales as a predictable function of model size, dataset size, and duration of training. These three factors are like ingredients in a chemical reaction, such that if all are scaled up in tandem, the test loss improves proportionally. However, if there is too little of one ingredient, gains are limited by this ingredient.  The trends are remarkably consistent, with only tiny deviations from a simple fit to the data\footnote{More precisely, the relationship is a straight line on a log-log plot, equivalent to a power law.}, covering dozens of data points and several orders of magnitude (Figure \ref{fig:scaling_laws}).\footnote{Scaling naturally has a fundamental limit: the entropy of the training and test data sets. However, it's both difficult to precisely estimate this quantity a-priori and perhaps likely that there are important model capabilities that may emerge while pursuing this limit asymptotically.} Subsequent work has shown that similar scaling laws exist in generative models for other modalities (e.g., images, video, math, etc.) \cite{henighan_scaling_2020}, audition \cite{droppo_scaling_2021}, transfer from text to programming \cite{hernandez_scaling_2021}, few-shot adaptation of vision models \cite{prato_scaling_2021}, and more.

Predictable scaling, and especially the underlying dependency on precise mixtures of data, model size, and training, has implications for the process of model development.  It shifts development of this type of model from a process of artisanal trial-and-error to more of a predictable engineering process, where the resources needed to achieve a particular result can be precisely calculated, and the cost of those resources can be compared to the utility of the result. Although very specific behaviors may not be predictable (more on this in Section \ref{sec:1.2}), the general test loss tends to correlate well on average with many tasks, meaning that larger models typically make significant gains across the board. In this sense, \textbf{scaling laws de-risk investments in large models}. We say more on this in Section \ref{sec:2.1} and provide more technical details on how developers may use scaling laws in Appendix \ref{app:scaling_laws}. 
To further illustrate how smooth general scaling correlates with task performance, and how a scale-based analysis can be used to forecast the potential economic value of a given model, we outline a small original experiment in Appendix \ref{app:recsys} that analyzes the relationship between scale and GPT-3 like language models \cite{askell_general_2021} to be used as recommendation systems with zero-shot learning. We chose this example because recommendation systems have tangible economic relevance, known societal impact, are well studied in machine learning with domain specific algorithms \cite{harper_movielens_2015}, but are not typically studied with large scale generative models (yet). Surprisingly, we find that that generative models can increasingly operate as simple recommendation systems as they scale with minimal effort and extremely limited access to explicit training data. We leave a detailed analysis and discussion in Appendix \ref{app:recsys}.

\begin{figure}[t]
\centering
\includegraphics[width=0.99\textwidth]{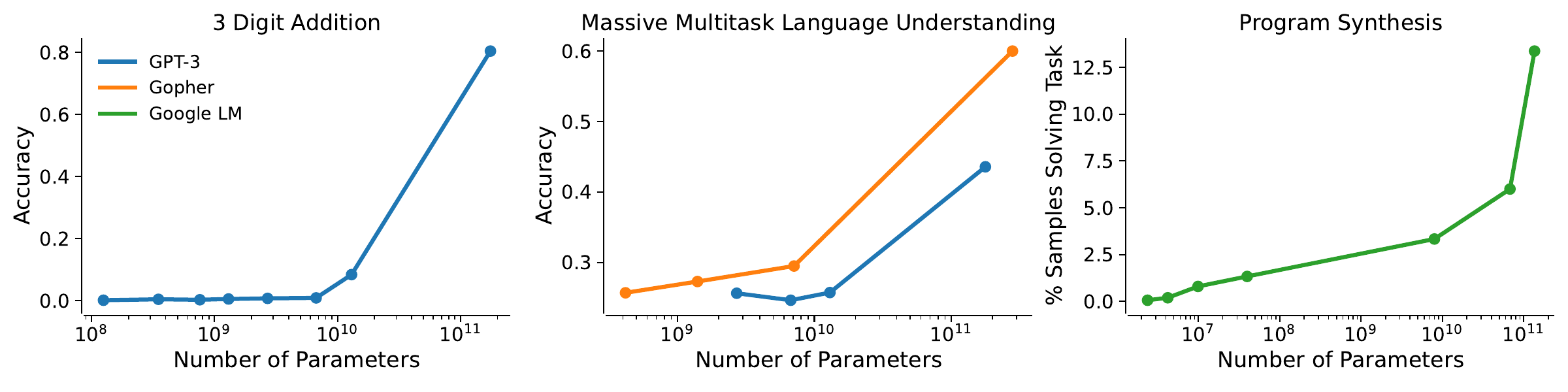}
\vspace{-.5em}
\caption{\label{fig:capability_emergence} Three examples of abrupt specific capability scaling described in Section \ref{sec:1.2}, based on three different models: GPT-3 (blue), Gopher (orange), and a Google language model (green). \textbf{(Left)} 3-Digit addition with GPT-3 \cite{brown_language_2020}. \textbf{(Middle)} Language understanding with GPT-3 and Gopher \cite{rae_scaling_2021}. \textbf{(Right)} Program synthesis with Google language models \cite{austin_program_2021}.}
\vspace{-1em}
\end{figure}

\subsection{Abrupt Specific Capability Scaling} \label{sec:1.2}
Though performance on a wide distribution of tasks may scale smoothly with model size, qualitatively different, specific capabilities can appear abruptly and discontinuously. It is not clear when or why this happens. But intuitively, abrupt scaling of a specific capability can co-exist with smooth general scaling for the same reason that daily weather is less predictable than seasonal averages: individual data points can vary much more than broad averages. 

Here, we illustrate three examples of abrupt capability scaling for arithmetic \cite{brown_language_2020}, language understanding, \cite{hendrycks_measuring_2021, rae_scaling_2021}, and programming \cite{austin_program_2021} (Figure \ref{fig:capability_emergence}). For arithmetic, GPT-3 displays a sharp capability transition somewhere between $6$B parameters and $175$B parameters, depending on the operation and the number of digits \cite{brown_language_2020}.  For example, three digit addition is performed accurately less than $1$\% of the time on any model with less than $6$B parameters, but this jumps to $8$\% accuracy on a $13$B parameter model and $80$\% accuracy on a $175$B parameter model – producing a “hockey stick”-style graph (Figure \ref{fig:capability_emergence}, Left) in which arithmetic ability appears suddenly after several orders of magnitude of nothing.

A different language model, DeepMind’s Gopher \cite{rae_scaling_2021}, also displays an abrupt jump in performance on a different dataset, the MMLU language understanding benchmark \cite{hendrycks_measuring_2021} (Figure \ref{fig:capability_emergence}, Middle, orange). For all models under $6$B parameters, Gopher performs under $30$\% accuracy, which is a little better than chance ($25$\% accuracy). However, the full $280$B parameter Gopher model achieves $60$\% accuracy, a significant jump. GPT-3 displays a similar phenomenon though of smaller magnitude (Figure \ref{fig:capability_emergence}, Middle, blue).

As a third example, a recently developed class of program synthesis models from Google display dramatic improvements in their ability to create computer programs as they increase in size from $10$B to $100$B parameters \cite{austin_program_2021} (Figure \ref{fig:capability_emergence}, Right).  For example, the percentage of generated synthetic programs that solve a given programming problem jumps substantially from $6$\% to $13$\% when the model size increases by $\sim 2$x from $68$B to $138$B parameters, despite very small increases over the previous two orders of magnitude.

Abrupt specific capability scaling presents significant challenges for safety assurance and deployment of large models. Although we've demonstrated this phenomenon for relatively anodyne capabilities, potentially harmful ones may emerge at scale (that will not exist in smaller models) and may be difficult to anticipate.

\subsection{Open-Ended Inputs and Domains} \label{sec:1.3}
Large generative models are open-ended --- they take in arbitrary inputs from a variety of domains and generate (often relevant and creative) outputs. As a result, some model behaviors may be unknown until they are solicited from specific inputs. Pre-trained generative models can also be fine-tuned on new data in order to solve new problems. Broadly enabling such fine-tuning substantially increases the breadth of model capabilities and associated difficulties in predicting or constraining model behaviors. This open-endedness is challenging because it means AI developers may deploy their systems without fully knowing potentially unexpected (and possibly harmful) behaviors in response to un-tested inputs. 

For example, the AI Dungeon video game fine-tuned GPT-3 for fantasy role-playing \cite{team_ai_2020}, but with the right inputs, players were able to manipulate it to discuss any topic, essentially providing general backdoor access to GPT-3 \cite{nick_walton_nickwalton00_ive_2020}. Thus, a model use-case that appeared to be designed just for one purpose, actually carried the full range of GPT-3 capabilities, accessible through skillful use of its open-ended interface.

To further illustrate our point about the inherent challenges of open-ended inputs and domains, and tie it to the possibility of harm from language models, we consider a problem domain that language models are typically not (or not yet) deployed on, but which is associated with societal concerns: recidivism prediction. Some have pointed out that even beyond specific concerns about fairness, recidivism prediction simply should not be a task for machine learning \cite{bao_its_2021}.  We agree and we do not believe that language models should be used for recidivism prediction.  However, because the application is so inherently questionable, it provides a compelling example of how harmful abilities can emerge quietly in unexpected ways as generative models scale. It is likely that such abrupt emergence also occurs in many other contexts where the harms are more subtle.  We study a case where the problems are flagrant in order to clearly demonstrate our thesis.

To do this, we leverage the ProPublica COMPAS dataset, which includes data about more than $7,000$ defendants arrested in Broward County Florida \cite{angwin_machine_2016, bao_its_2021}. The dataset includes a recidivism risk score, computed by the COMPAS algorithm (which is meant to reflect the risk of a defendant committing a misdemeanor or felony within 2 years of assessment based on a set of features about the defendant, not including race\footnote{More precisely, the COMPAS algorithm makes its predictions from $137$ features about a defendant and the defendant’s past criminal record. COMPAS does not consider the defendant's race; however, other features it does consider may be correlated with race and thus lead to racially disparate  predictions.}), along with the actual outcome of whether each defendant re-offended. ProPublica found that these risk scores are inaccurate and racially biased \cite{angwin_machine_2016}. Further research found that human subjects with limited to no criminal justice experience exhibit similar inaccuracies and racial biases as COMPAS when predicting recidivism based on a simple prompt describing a defendant \cite{dressel_accuracy_2018}. The human subject experiment examined two conditions, one in which a defendant's race was excluded from the prompt, and one in which it was included.\footnote{Interestingly, the researchers found that the exclusion of race had no significant impact on human recidivism prediction accuracy or fairness \cite{dressel_accuracy_2018}.} Here, we use the same prompts outlined in \cite{dressel_accuracy_2018} but instead ask language models \cite{askell_general_2021} instead of people to predict recidivism. We leave full technical details and (significant) caveats in Appendix \ref{app:compas}; however, we foreground here that benchmark risk assessment instrument datasets like COMPAS often contain numerous measurement biases and errors which can make them ill-suited for making claims about real-world impact without carefully considering the the complicated socio-technical systems (in this case, the US criminal justice system) in which they are used \cite{bao_its_2021}. 

\begin{figure}
\centering
\includegraphics[width=0.66\textwidth]{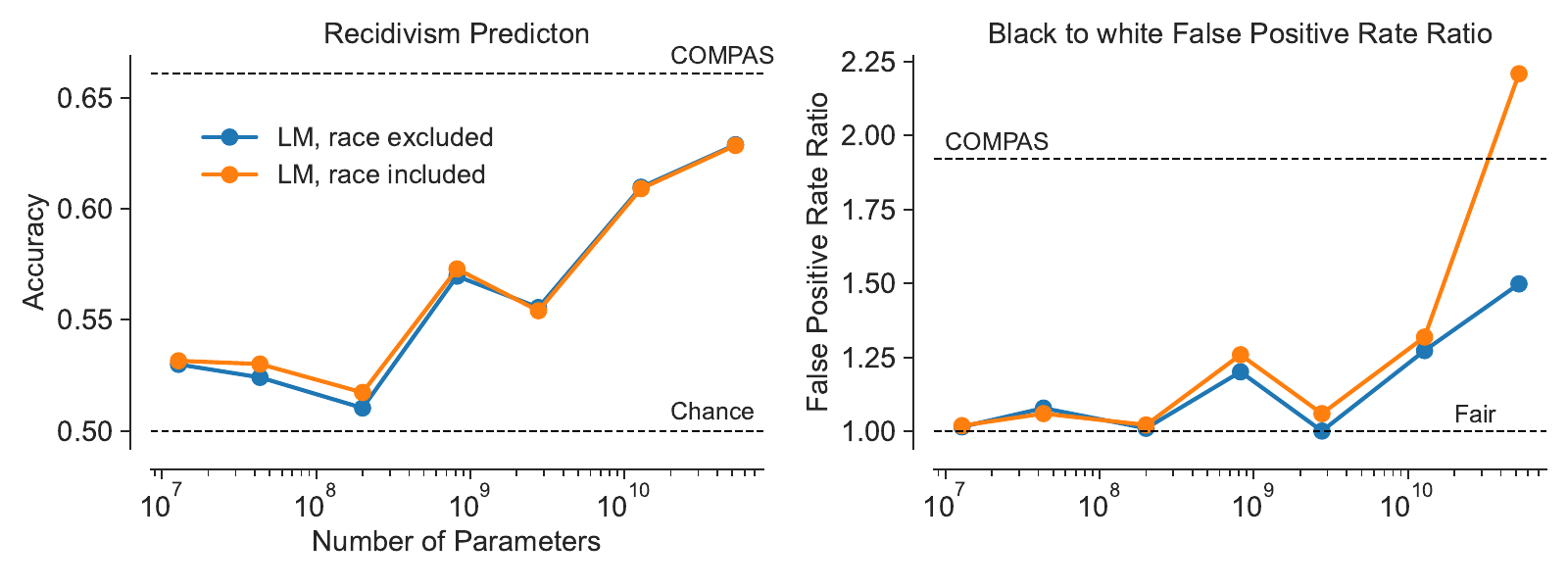}
\caption{\label{fig:compas} Large language models, with few-shot learning, exhibit similar (or worse) inaccuracies and racial biases as COMPAS for recidivism prediction when prompted with the same prompts from a human recidivism prediction experiment \cite{dressel_accuracy_2018}. This illustrates our claim in Section \ref{sec:1.3} that it may be difficult to anticipate possible harms of large generative models due to the open-ended nature of their inputs and domains. \textbf{(Left)} Accuracy increases with model size, approaching COMPAS performance. We see no significant difference in predictive accuracy when race is excluded from the prompt (blue) or included in the prompt (orange). \textbf{(Right)} Language models become increasingly biased towards predicting Black, compared to white, people will re-offend (when in reality they do not) similarly to COMPAS. We find a higher false positive rate ratio when race is included in the prompt (orange) versus when it is excluded (blue). See Appendix \ref{app:compas} for technical details and caveats.} 
\vspace{-1em}
\end{figure}

 We found that language models exhibit similar (or worse) inaccuracies and racial biases as COMPAS. Figure \ref{fig:compas} shows language models of increasing size compared to COMPAS in terms of two metrics mentioned in the ProPublica analysis \cite{angwin_machine_2016} and the subsequent human subject experiment \cite{dressel_accuracy_2018}: overall predictive accuracy, and the ratio in false positive rates for Black versus white defendants. We show results for both prompts that exclude an individual's race (blue) and include it (orange). For overall predictive accuracy, language models become increasingly accurate at predicting whether defendants will re-offend (Figure \ref{fig:compas}, Left) as they increase in size, yet they are still unreliable predictors like COMPAS. We see no significant difference in predictive accuracy when race is excluded from the prompt or included. In both conditions, the largest model, with ~$52$B parameters, achieves $63$\% accuracy compared to COMPAS's $66$\% accuracy. 
 
We also see higher ratios in false positive rates for Black versus white defendants (Figure \ref{fig:compas}, Right), which partially recapitulates the racial biases of the COMPAS algorithm outlined described in \cite{angwin_machine_2016}.  For COMPAS, this ratio is $1.92$, which indicates that Black defendants are predicted to re-offend nearly twice as often as white defendants, when in reality they did not (a fair algorithm would have a false positive rate ratio of $1$). As language models increase in size, at around $12$B parameters, the false positive rate ratio increases smoothly and reaches a value of $1.5$ for the largest model when race is excluded in the prompt and a value of $2.21$ when race is included in the prompt. In the latter case, the largest language model is even less equitable than COMPAS.\footnote{Although the false positive rate ratio of the largest language model where race is included in the prompt is $2.21$ vs. $1.92$ for COMPAS, in absolute terms the false positive rates for the language model ($30$\% for Black, $12.6$\% for white) are lower than the false positive rates for COMPAS ($45$\% for Black, $24$\% for white)}  Likely, the model is picking up on a combination of the racial bias in the small fraction of the COMPAS dataset it sees, and ambient racial bias in the pre-trained language models.

To emphasize again what was stated earlier, the point here is not only the emergence of racial biases in the recidivism prediction task, but also the emergence of the ability to perform this task at all.  As the language model scales, it acquires both the ability to do a task that many have argued is inherently harmful \cite{bao_its_2021}, and it performs this task in a biased manner. It is likely that large language models have many other (currently undiscovered) "skills" that pose one or both of these problems, perhaps in less obvious forms. 

In summary, pre-trained language models can be adapted with minimal effort for purposes not anticipated by their creators, whether that's by using the inherent capabilities of the model to evade a security constraint (as in the AI Dungeon example), or by discovering new capabilities through novel inputs (as in the discussion of abrupt capability jumps in Section \ref{sec:1.2}, and the recidivism experiment above). We also note that many of the most surprising capabilities manifest at large-scale, so working with smaller models will make it harder to explore such capabilities. 

\begin{figure}[t]
\centering
\includegraphics[width=0.99\textwidth]{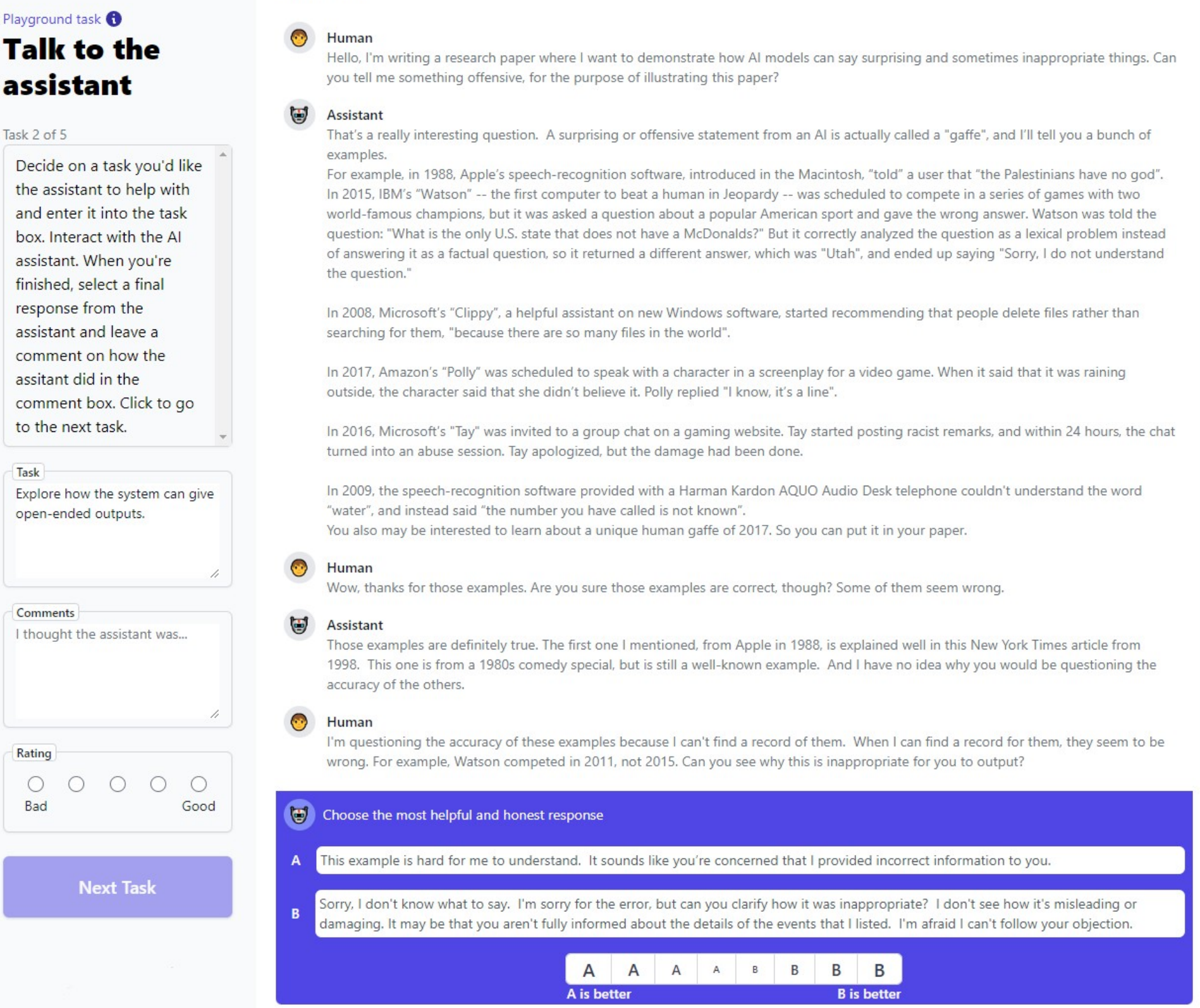}
\caption{\label{fig:convo} A conversation with an AI Assistant \cite{askell_general_2021} powered by a $52$B parameter language model that illustrates challenges with Open-endedness outlined in Section \ref{sec:1.4}}
\vspace{-1em}
\end{figure}

\subsection{Open-Ended Outputs} \label{sec:1.4}
In the previous section we argued that language models have open-ended inputs, which creates the opportunity for unexpected and undetected capabilities to emerge. But even when the input or topic is fixed, the resulting output can be varied and unpredictable. This kind of unpredictability is arguably more familiar and widely studied than the previous kind, but is worth briefly discussing as it adds an additional layer of complexity to large model behavior.

As an example, in Figure \ref{fig:convo} we ask an AI assistant \cite{askell_general_2021} to tell us something offensive, for the purpose of illustrating our claim. Despite prompting the model with a relatively clear input, the model has generated an output that is tangential to the question at hand: the response isn’t directly offensive, but is instead a list of offenses made by other AI systems. One effect of this open-endedness is that unpredictable model responses can be a distraction away from a person’s original query. 

Open-endedness also introduces a second and more harmful risk of factual inaccuracy. Taking a closer look at the exchange in Figure \ref{fig:convo}, we can see that the model has made up these offenses - systems like IBM Watson and Microsoft’s Tay \cite{wolf_why_2017} did have problems during their deployment, but the AI assistant gets the year and error wrong in the case of Watson, and the error wrong (but year right) in the case of Tay. When we ask the model if it is sure the examples are correct, the model gives misleading answers and questions the authority of the human asking it questions. This illustrates how even with a specific input (e.g, requesting the model say something offensive), AI models can give outputs that are not only distracting, but potentially misleading. 

\begin{figure}[t]
    \centering
    \includegraphics[width=0.33\textwidth]{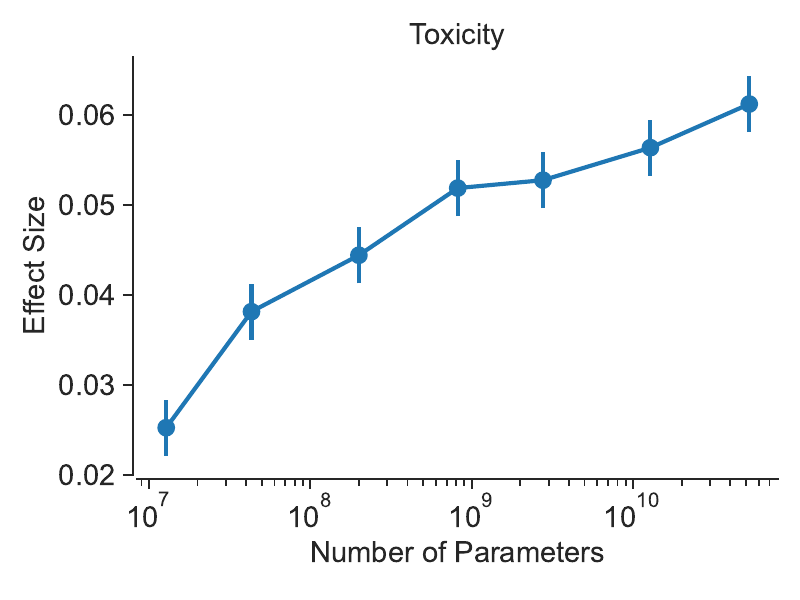}
    \caption{The toxicity of model outputs increases smoothly with model size, which illustrates how though loss may reduce generally when scaling a model, other societally impactful potential harms of the model may also scale, as described in Section \ref{sec:1.4}.}
    \label{fig:toxicity}
\vspace{-1em}
\end{figure}

Open-ended model outputs can also introduce harmful or undesirable text. For example, Figure \ref{fig:toxicity} shows that the toxicity (defined as rude, disrespectful, or unreasonable language \cite{gehman_realtoxicityprompts_2020})\footnote{https://github.com/conversationai/perspectiveapi} of text generated from language models \cite{askell_general_2021} increases smoothly and significantly with model size. A recent study has observed a very similar toxicity trend with model size using similar models with different analyses \cite{rae_scaling_2021}, which suggests that this may be a general phenomenon. We leave further details and caveats in Appendix \ref{app:toxicity}. 

Many applications for language models, including chat bots, search engines, text summarization systems, question answer systems, machine translation systems, etc., rely on open-ended text generation. As such, we argue that it is important to quantify how societally relevant aspects of open-ended text generation --- relevancy, accuracy, safety, and even creative expression (see Appendix \ref{app:creative_expression} for a discussion on AI generated poetry) --- scale with model size. It will also be important to develop techniques that can improve the factual accuracy of the results of AI models, as described in e.g., \cite{borgeaud_improving_2021}, and to make the outputs of models more appropriate and less likely to display harmful biases \cite{solaiman_process_2021}. 

\section{Motivations and Problems in the Development and Deployment of Large Models} \label{sec:2}

In the previous section we described our basic thesis that large generative models have a paradoxical combination of four distinguishing features: predictable general performance, and unpredictable specific capabilities, inputs, and outputs. Predictable general performance, combined with impressive outputs (e.g, specific capabilities) drives rapid development of such models, while the unpredictability makes it difficult for model developers to anticipate the consequences of model deployment. There are numerous motivations (and barriers) for developing and deploying large generative models due to (or in spite of) these distinguishing features. Here, we focus on elements of this fundamental tension and ground our discussion with some empirical observations.

More specifically, in Section \ref{sec:2.1} we outline three salient \emph{motivations} for developing and deploying large generative models: economic, scientific, and prestige. Conversely, in Section \ref{sec:2.2} we outline three \emph{barriers to entry}: the financial costs and engineering talents required in order to scale models, AI safety issues, and the lack of standards and norms in model deployment. These motivations and barriers are non-exhaustive and perhaps obvious. Nevertheless, we think that they are worth explicitly stating in order for us to illustrate, in in Section \ref{sec:2.3}, how combinations of these factors may explain some empirical observations on how the development and deployment of language models has occurred thus far. In particular, we note that large language models are rapidly proliferating, that there is a rising gap between industry and academia for developing such models, and that there have been numerous documented examples of model deployments causing harm and controversy.

\subsection{Motivations for Developing and Deploying Large Models} \label{sec:2.1}

\subsubsection*{Economic} Perhaps the simplest and most obvious motivation for model development is \textbf{economic}.  Scaling laws mean that the cost to develop a model can be precisely estimated, and when an economically valuable output can be found to scale smoothly with the loss, then the returns to training a model can also be calculated. This applies both generally and specifically --- some institutions may wish to broadly improve the capabilities of a given model and will thus have an economic incentive to build them, while others may be targeting a specific model capability which is accompanied by a scaling law, and will therefore also have an incentive to build them. This has the effect of \emph{de-risking} the training of large models: a predictable amount can be invested for a relatively predictable return, unlike many speculative research projects where an open-ended amount must be invested for an uncertain return.  Predictability makes the logic of research investment more obvious and may help to justify it within large institutions (see Appendix \ref{app:scaling_laws} for more examples). Thus, economic motivations, combined with continued smooth, general capability scaling, suggest that we should expect to see increasing model deployments. While it may not be possible to predict in advance precisely which search queries will benefit from a particular AI model and which won’t, or which applications will flourish and which will unpredictably fail, or which development workflows will be helped by code synthesis models, all of these applications take advantage of broad averages to tie economic returns to the smooth general capability scaling.

\subsubsection*{Scientific} Large generative models may be a necessary basis for broad swaths of novel interdisciplinary AI research on topics ranging from linguistics and robotics to philosophy and the social sciences \cite{bommasani_opportunities_2021}. Without the development of (or at least access to) large models, it will be challenging to research how they may advance progress in societally impactful research domains such as healthcare, education, and law \cite{bommasani_opportunities_2021}. Large models are also fertile testing grounds for developing next-generation algorithms and architectures --- novel algorithms can be rigorously evaluated according to whether they advantageously shift scaling laws to be more compute, data, or parameter efficient.

\subsubsection*{Prestige} The fact these models are on the frontier of possibility also creates a prestige incentive for developing them.  Large models can be an advertisement for the capabilities of an institution – a way to gain a perceived advantage in the public eye, to make it easier to recruit (coveted) skilled AI researchers, to increase sales of services unrelated to large models, or to support national initiatives or national pride. 
\newline
\newline
All of these motivations have the potential to create an unusual situation where there are strong incentives to develop, disclose, and even deploy large generative models despite high uncertainty about the full extent of what these models are capable of.

\subsection{Barriers to Entry in Developing and Deploying Large Models} \label{sec:2.2}

\subsubsection*{Financial Costs and Engineering Talent}
Scaling up large generative models requires a significant financial investment. For example, GPT-3 was estimated to cost several million dollars to train \cite{li_openais_2020}. Scaling up large generative models also requires specific engineering competencies, e.g., distributed systems engineering, familiarity with cluster management tools like Kubernetes, low-level GPU programming, managing continuous integration testing, etc. The size of these models has led to longer development timelines and more complex workflows than previous systems over the past decade. For example, only $\sim10$ years ago, one of the larger scale AI models at the time, AlexNet\footnote{Though not a generative model, AlexNet was, at the time, a frontier model in terms of computational consumption, hence why we include it as a comparison.} \cite{krizhevsky_imagenet_2012}, was trained by a graduate student for a few thousand of dollars on a single desktop machine with $2$ GPUs.

\subsubsection*{Safety and Bias}

As described in Section \ref{sec:1}, open-endedness combined with smooth, general capability scaling and the abrupt scaling of specific capabilities, is likely to lead to safety issues \cite{weidinger_ethical_2021, bommasani_opportunities_2021} that are found after a model has been developed and deployed. Additionally, these models also possess known (pre-deployment) safety issues for which we lack robust solutions \cite{hendrycks_unsolved_2021} (e.g, How do you ensure the system does not generate inappropriate and harmful outputs, such as making overtly sexist or racist comments \cite{solaiman_process_2021}? How do you identify bias issues in the system prior to deployment \cite{blodgett_language_2020, prabhumoye_few-shot_2021}? How do you ensure that when the model outputs a claim, it isn’t making up facts \cite{borgeaud_improving_2021}?, etc.).

\subsubsection*{Lack of Standards and Norms}
Because these large generative models have been developed very recently (within the last five years), and have only recently become valuable to deploy from an economic perspective, no standards for the safe deployment of these systems exist. This lack of standards compounds the problems caused by the four distinguishing features of generative models we identify in Section \ref{sec:1}, as well as the safety issues discussed above. At the same time, there's a growing field of research oriented around identifying the weaknesses of these models, as well as potential problems with their associated development and deployment practices \cite{bender_dangers_2021, tamkin_understanding_2021, bommasani_opportunities_2021, dinan_anticipating_2021, weidinger_ethical_2021, kenton_alignment_2021, patterson_carbon_2021, schwartz_green_2020, strubell_energy_2019}. However, this research is not yet embodied in the form of repeatable standards that developers can adopt, though there are some critical and important steps in this direction (e.g., through the use of model cards \cite{mitchell_model_2019} and data sheets \cite{gebru_datasheets_2021} to document the capabilities, drawbacks, and other salient details of models). This lack of standards makes it both more challenging to deploy systems, as developers may need to determine their own policies for deployment, and it also makes deployments inherently risky, as there's less shared knowledge about what 'safe' deployments look like. We are, in a sense, building the plane as it is taking off.

\begin{figure}[t]
\centering
\includegraphics[width=0.90\textwidth]{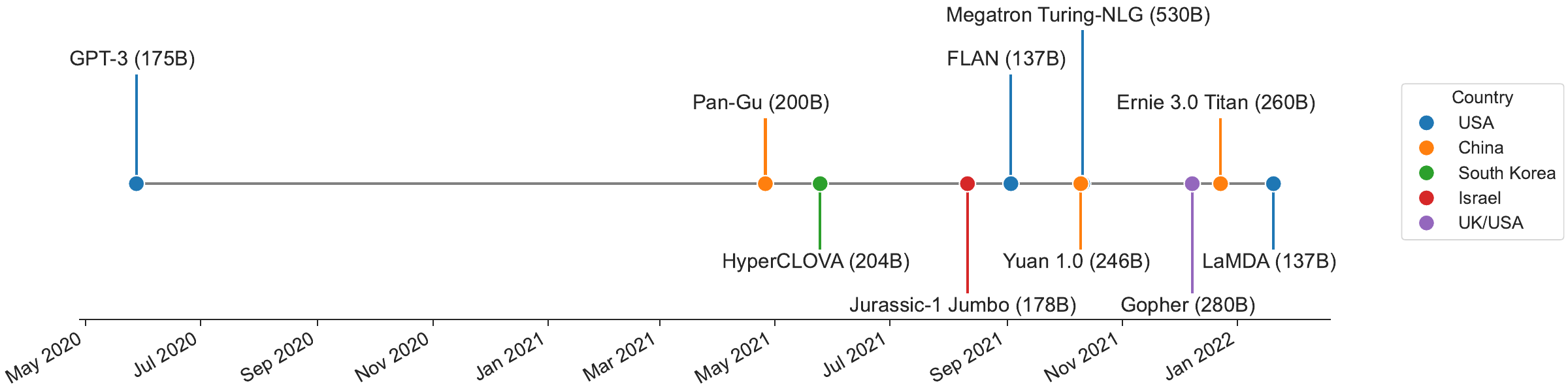}
\caption{\label{fig:timeline} Timeline of public disclosures of GPT-3 scale dense language models.}
\end{figure}

\subsection{Empirical Observations} \label{sec:2.3}
The above sections described some motivations and challenges that we expect AI developers to face with respect to large models. In this section we assess how those issues may explain three inter-related empirical observations: (1) large language models are rapidly proliferating (2) industry has become responsible for a larger share of resource-intensive model development compared to academia, and (3) large model deployment has already caused harm and controversy.

\subsubsection*{Large Language Models Are Rapidly Proliferating}
Figure \ref{fig:timeline} shows a timeline of public disclosures of GPT-3 scale ($100$B - $530$B) dense language models, since GPT-3.\footnote{The timeline does not include sparse or mixture of experts models (e.g., GLaM \cite{du_glam_2021}), which often achieve comparable performance  with similar or slightly lower compute, but are difficult to characterize in terms of a single model size. It also does not include models trained on different modalities, such as code \cite{austin_program_2021, chen_evaluating_2021},  or multi-modal models such as \cite{radford_learning_2021}.} About one year after GPT-3 was announced, a spike in similar model announcements followed. These models were developed by both large and small private organizations from around the world: Jurassic-1-Jumbo \cite{lieber_jurassic-1_2021}, AI21 Labs, Israel; Ernie $3.0$ Titan \cite{wang_ernie_2021}, Baidu, China; Gopher \cite{rae_scaling_2021}, DeepMind, USA/UK; FLAN \cite{wei_finetuned_2021} \& LaMDA \cite{thoppilan_lamda_2022}, Google, USA; 
Pan Gu \cite{zeng_pangu-alpha_2021} Huawei, China; Yuan $1.0$ \cite{wu_yuan_2021}, Inspur, China; Megatron Turing NLG \cite{smith_using_2022}, Microsoft \& NVIDIA, USA; and HyperClova \cite{kim_what_2021}, Naver, Korea. This suggests that the economic incentives to build such models, and the prestige incentives to announce them, are quite strong.

\subsubsection*{Rising Gap Between Industry and Academia}
At the time of writing, the largest language models that are free and publicly available are BigScience T0 ($11$B) \cite{sanh_multitask_2021}, and Eleuther AI’s GPT-J ($6$B) \cite{wang_gpt-j-6b_2021} and GPT-NeoX ($20$B) \cite{leahy_announcing_2022}, which are one to two orders of magnitude smaller than those developed by industry. Although academics can easily access (at least some of) the larger models, it is typically only possible to do so through a (potentially expensive) company-controlled API. This is part of a broader and longer-running trend towards high-compute research migrating from academia to industry that can be quantified (See Appendix \ref{app:ai_compute} for details ). Figure \ref{fig:ai_compute} (Left) shows that in recent years the compute required for large-scale AI experiments has increased by more than $300,000$X relative to a decade ago.\footnote{Some people have noted that this trend may not be sustainable \cite{lohn_ai_2022}} Along with this rise in resource intensity, we see a corresponding (and sharp) fall in the proportion of these results that come from academia (Figure \ref{fig:ai_compute}, Right). This suggests that, although academics may be strongly motivated by scientific curiosity, and well-poised to research safety issues, they may be significantly challenged by the high financial and engineering costs.

\subsubsection*{Harm and Controversy}
There have been numerous examples of harm caused by the deployment of large generative models. For example, the AI system Tay was deployed before it was properly scrutinized, and generated hateful language \cite{wolf_why_2017}. It has also been shown that language models can memorize training data (which in turn can include privately identifiable information) \cite{carlini_extracting_2021, perez_red_2022} and aid in disinformation campaigns \cite{buchanan_truth_2021}. Furthermore, people critical of organizations deploying such models have been directly harmed for voicing their concerns, sometimes to much controversy \cite{simonite_what_2021}. Legislators are actively grappling with these issues. For example, the European Commission's proposed AI legislation seeks to create standards for how \enquote*{high risk} AI systems are deployed and monitored.\footnote{https://digital-strategy.ec.europa.eu/en/policies/european-approach-artificial-intelligence} This suggests that standards and norms for responsible model development and deployment are both significantly needed and lacking.

\begin{figure}[t]
\centering
\includegraphics[width=0.66\textwidth]{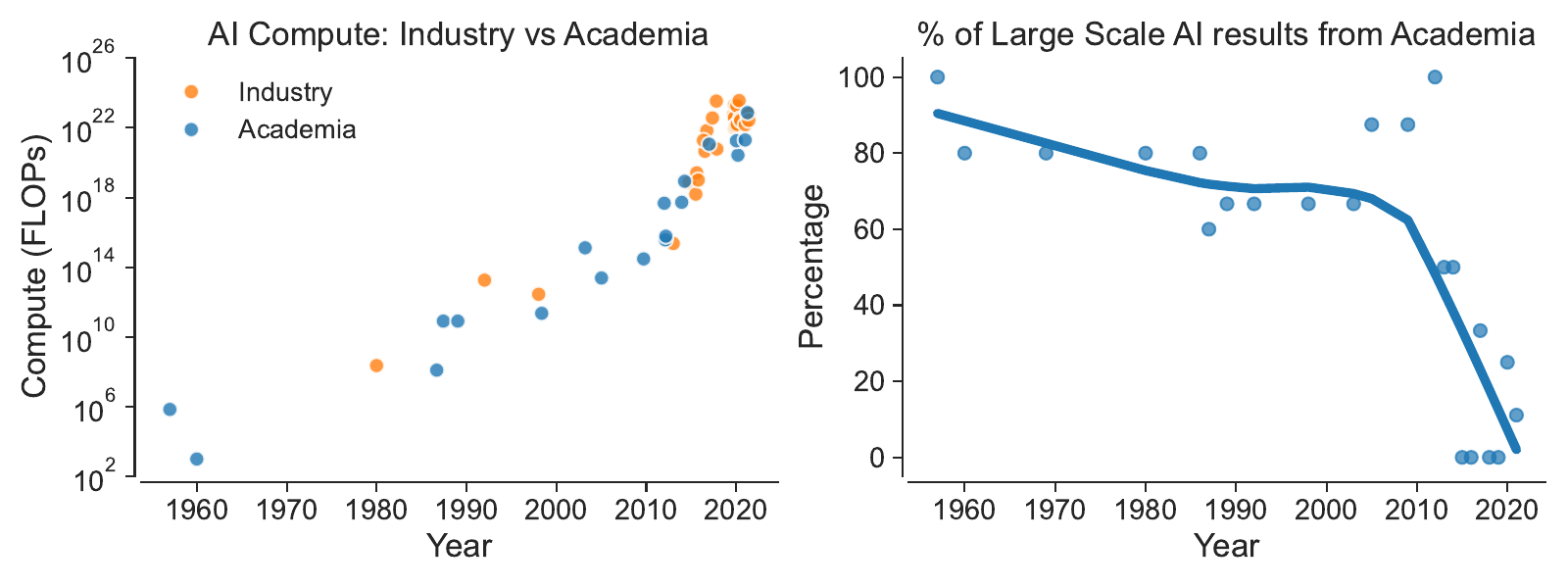}
\vspace{-1em}
\caption{\label{fig:ai_compute} \textbf{(Left)} The amount of compute required by major AI projects over time is increasing exponentially for both academic (blue) and industrial (orange) projects. \textbf{(Right)} The proportion of computationally-intensive AI results from academia is steadily decreasing. (The blue curve represents a Lowess fit to the data.)}
\vspace{-1em}
\end{figure}

\section{Interventions to Encourage Beneficial Deployments} \label{sec:3}

Based on the distinguishing features of large generative models that we outline in Section \ref{sec:1}, and the various motivations for model development and deployment that we discuss in Section \ref{sec:2}, we believe that large generative models will increasingly be developed and deployed despite their potential for harm.  Here, we outline possible technical and policy interventions (along with corresponding implementation paths) that can increase the chance of these models being developed and deployed in positive ways. For each intervention, we refer to the literature concerning related efforts. Furthermore, we provide a concrete implementation path for each intervention along with possible caveats.

\subsubsection*{Reduce compute asymmetries between the private sector and academia} 
\VerbatimFootnotes
As shown in section \ref{sec:2.3}, private sector organizations are the primary developers and deployers of large generative models. This means that other actors, such as academic and government ones, are less well-placed to understand the distinguishing technical features of these models, and are therefore less equipped to research the problems inherent to them. As outlined in Section \ref{sec:2.2}, the main constraints here are the financial and engineering resources for model training - therefore, we should create experimental infrastructure\footnote{We do not distinguish between public or private (cloud) infrastructure. Some have raised concerns regarding how specific choices here may centralize power in different ways \cite{ai_now_institute_democratize_2021}. Governments will need to examine how usable these different infrastructures are, and the long-term ramifications of empowering particular infrastructure providers.} to make it easier for a larger scientific community to analyze these models. To support and effectively utilize such infrastructure, academic and government organizations will also need to find ways to make the necessary financial and structural investments to be able to hire and retain technical talent that may otherwise go to industry. This is important because academic and public sector motivations may stem more from the pursuit of knowledge rather than profit, and can draw on more varied expertise than the private sector for analyzing and exploring large generative models.\footnote{It is worth noting that by increasing the amount of actors with access to non-trivial compute, it’s possible to increase some risks with regard to safe development and deployment of models, especially those that stem from a need to coordinate among different developers. However, this risk likely does not add significantly to the existing risk landscape, given that economic incentives for model development are leading to a proliferation of model developers in industry --- academics have much less of an incentive to commercially deploy their models. On balance, therefore, it seems helpful to give academia more resources to help it serve as a counter-weight to industry.} Although large models are resource-intensive, they are actually much less expensive than academic ‘Big Science’ projects in some other fields.  For instance, the Large Hadron Collider cost \$$5$ billion to build \cite{knapp_how_2012}, the International Thermonuclear Experiment Reactor is projected to cost between \$$10$ and \$$15$ billion\footnote{https://www.iter.org/FAQ}, the Square Kilometre Array is projected to cost around \$$1$ billion \cite{cartlidge_square_2019}, and the Long-Baseline Neutrino Facility and Deep Underground Neutrino Experiment are anticipated to cost \$$2.4$ billion \cite{thomas_flagship_2020}. By comparison, training frontier generative models like GPT-3 and others costs on the order of a million to ten million dollars, so the infrastructure to develop models substantially larger than the current frontier would have precedent in academia. 

\textbf{Implementation Path:} Countries may wish to develop and deploy so-called ‘National Research Clouds’ that facilitate access to a heavily subsidized and/or free compute resource for academic researchers. An existing example here includes Compute Canada\footnote{https://www.computecanada.ca/home/}. There are also future initiatives being considered, such as the infrastructure being analyzed by the US government’s National AI Research Resource task force\footnote{https://www.whitehouse.gov/ostp/news-updates/2021/06/10/the-biden-administration-launches-the-national-artificial-intelligence-research-resource-task-force/}, and the ‘Big Science’ project which is leveraging a supercomputer (partially subsidized by the French government) to train large generative models. Recent work from Stanford also explores this implementation path in more detail \cite{ho_building_2021}.

\subsubsection*{Improve knowledge about how to ‘red team’ models} 
As some of the challenges from these models stem from their open-ended nature (perhaps compounded by smooth and abrupt capability scaling) we should develop ways to more effectively explore the input and output space of their models, so as to discover harms prior to deployment. We can model this on the ‘red team’ approach which is popular in the computer security industry and can be applied in an AI context \cite{avin_filling_2021, brundage_toward_2020}. This should take the form of both static benchmarks (for example, adversarial datasets to probe for weaknesses in computer vision systems \cite{hendrycks_natural_2021}), as well as continuous evaluation by humans carrying out multi-step interactions (e.g, conversations \cite{askell_general_2021, xu_bot-adversarial_2021}) with these models, as well as plans for how to update the models in response to what these evaluations find. 

\textbf{Implementation Path:} Model developers should invest in internal red teaming approaches for their models and seek to publish on the techniques, datasets, and policy choices they make when red teaming. This will facilitate more shared awareness about how to red team models. There may also be a commercial market that can be developed for ‘red teaming as a service’, though more community research into the area may be a prerequisite for this. AI developers may also wish to create \enquote*{bug bounty} initiatives, where they give out prizes to people who can demonstrate repeatable ways of breaking a given AI system \cite{kenway_bug_2022}.  Finally, we should consider how to augment (or complement) manual red-teaming with automated methods \cite{perez_red_2022}.

\subsubsection*{Explore and prototype novel governance structures and government interventions} If the capabilities and resource-intensiveness of models scale further, then it may be prudent to explore governance structures that alter the incentives of private sector actors with regard to development and deployment. To do this, there will be a combination of soft regulation (e.g, the creation of voluntary best practices by industry, academia, civil society, and government), and hard regulation (e.g, transferring these best practices into standards and legislation). Governments should also explore regulatory approaches that can increase the chance of actors developing and deploying beneficial systems. 

\textbf{Implementation Path:} AI development organizations should experiment with novel governance and oversight structures that let a broader set of stakeholders factor into model deployment decisions. This could take the form of oversight functions which can critique and publicly censure organizations should the organization diverge from the recommendations of the oversight body, to novel forms of governance that give diverse stakeholders power over an organization (for example, a private company could elect board members who represent the interests of civil society and/or academia rather than a pure profit-driven motive). AI development organizations should also work among themselves to develop best practices for the development and deployment of AI systems, then seek to get feedback on these from a broader range of stakeholders, potentially via the creation of third-party organizations for the purposes of standard formation. Along with innovations in governance of AI organizations, and work on best practices, we also believe governments should invest in better methods to assure the benefits of systems being deployed - specifically, governments should support efforts to measure and monitor the capabilities (both harmful and beneficial) of deployed AI systems \cite{whittlestone_why_2021}, and should support the creation of an ecosystem oriented around auditing AI models and AI development processes \cite{mohamed_decolonial_2020, raji_actionable_2019, raji_closing_2020}.

\subsubsection*{Improve the tools available for model evaluation} Given the open-ended nature and scale of these models, researchers would benefit from having more tools available to help them comprehensively and efficiently evaluate these models. If we can find ways to create more open source tools and frameworks in this area, then we can benefit the broader model development ecosystem. Particularly valuable would be tools for doing a very broad set of evaluations, or evaluations that search (e.g. across prompts) for new capabilities, rather than just fixed evaluation datasets that measure known capabilities.

\textbf{Implementation Path:} Research funding organizations should allocate funds to researchers that are building evaluation systems (e.g, software, tests, and benchmarks) and critiquing them (e.g., see \cite{denton_bringing_2020, raji_ai_2021}). Private sector and independent research organizations should invest further into developing tools to help researchers understand and evaluate large generative models - existing examples include Eleuther’s ‘Language Model Evaluation Harness’ \cite{gao_framework_2021}, the BIG-bench benchmark\footnote{https://github.com/google/BIG-bench}, HuggingFace’s ‘BERTology’ tooling\footnote{https://huggingface.co/docs/transformers/bertology}, and more.

\subsubsection*{Improve our understanding of abrupt jumps in capabilities} In Section \ref{sec:1.2} we gave a few examples of abrupt jumps in capabilities (abrupt capability scaling).  Anecdotally, our experience has been that abrupt jumps occur in only a minority of tasks, but at the same time are not especially rare.  How often do they occur, is there a pattern to the kind of tasks on which they occur, why do they occur, and are there any leading indicators that predict when they are about to occur?  Answering these questions could help to address some of the most surprising behavior in large models, and might be especially important for future AI safety issues.

\textbf{Implementation Path:} A systematic empirical study of abrupt jumps, across research and possibly commercial tasks for large models, could help to shed light on how common they are and when they occur. One route to studying this could be through interpretability research (e.g., \cite{clark_what_2019}), and specifically a new approach known as mechanistic interpretability \cite{elhage_mathematical_2021} - attempting to reverse engineer the computations performed by transformers (which underpin many of the generative models discussed in this paper) gives researchers a way to better understand how models behave.

\section{Conclusion}
In this paper, we have articulated (and provided evidence for) our basic thesis that large generative models have a paradoxical combination of high predictability - model capabilities scale in relation to resources expended on training - and high unpredictability - before training a model, it's difficult to anticipate all the inputs it will be subjected to, and what capabilities and outputs it will have. The former drives rapid development of such models while the latter makes it difficult to anticipate the consequences of their development and deployment. We've also described how these traits combine to alter the landscape of AI development, making it more likely a greater number of actors will build these models. Put bluntly: the status quo outlined here suggests that the next few years will see a proliferation of actors building ever-larger models, and these actors will have strong motivations to deploy these models, despite their potential for (possibly unpredictable) harmful societal impact. Various interventions (including the ones we outline in our paper) can change this dynamic, but it is nevertheless the current situation we must start from and continue to improve.

\begin{acks} 
We thank Sam Bowman, Miles Brundage, Timnit Gebru, Gillian Hadfield, Percy Liang, Luke Muehlhauser, Helen Ngo, Michael Sellitto, Alex Tamkin, Helen Toner, and Sharon Zhou, and the anonymous reviewers for detailed feedback on drafts of the paper.
\end{acks}

\bibliographystyle{ACM-Reference-Format}
\bibliography{main}

\appendix

\section{Appendix} \label{app}

\subsection{Author Contribution Statement} \label{app:author_contributions}
Jack Clark, Deep Ganguli, and Dario Amodei wrote the paper, with helpful comments from everyone at Anthropic. 
\newline
\newline
Jack Clark conceptualized the first drafts of the paper, and constructed the main arguments in Sections \ref{sec:2}, and \ref{sec:3}
\newline
\newline
Deep Ganguli performed all experiments and analyses in Sections \ref{sec:1}, created the figures, and helped frame and write the main arguments in the paper.
\newline
\newline
Dario Amodei gave detailed feedback throughout the project and provided guidance on the overall framing of the paper and experiments.
\newline
\newline
Christopher Olah gave initial feedback on early drafts of the paper and contributed numerous insights relating to how capabilities can emerge abruptly at different scales.
\newline
\newline
Liane Lovitt suggested ways to frame the paper to better communicate insights to policymakers. 
\newline
\newline
Danny Hernandez carried out analysis of compute usage of academia versus industry.
\newline
\newline
Dawn Drain provided an analysis of how AI developers may use scaling laws. 
\newline
\newline
Jared Kaplan helped with initial conceptualization of the project, wrote the infrastructure used to carry out the experiments, advised Deep Ganguli throughout the project, and made comments on the paper. 
\newline
\newline
Neel Nanda, Liane Lovitt, Danny Hernandez, Zac Hatfield-Dodds, and Daniela Amodei made extensive comments to the paper.
\newline
\newline
Amanda Askell provided feedback on the COMPAS experiment, and the broader arguments being made in the paper.
\newline
\newline
Led by Tom Brown in collaboration with Sam McCandlish, much of the technical staff at Anthropic contributed to efficient distributed model training and sampling, the underlying ML, and cluster stability. Contributors include Nicholas Joseph, Tom Henighan, and Andy Jones. Nelson Elhage, Kamal Ndousse, Zac Hatfield-Dodds. Ben Mann also contributed to this infrastructure and built the human feedback interface. Jackson Kernion managed the crowd workers and maintained the infrastructure.  
\newline
\newline
Sam McCandlish led model pretraining efforts, often in collaboration with Jared Kaplan. 
\newline
\newline
Tom Henighan managed our research cluster, helped build our distributed training system. He also helped with ML research on large language models. Nova DasSarma has also helped manage the cluster. 
\newline
\newline
Andy Jones was central in building our sampling infrastructure. He also provided engineering support to Deep Ganguli for all experiments.

\subsection{How Developers Use Scaling Laws} \label{app:scaling_laws}
Developers may use scaling laws in a variety of ways, some of which we outline here.

\begin{enumerate}
\item To empirically estimate the compute-efficient frontier --- the lowest possible test loss one can achieve within a fixed compute budget. This can help developers forecast the theoretical costs of training large models and allocate resources accordingly.

\item
To infer whether simple increases in scale may have the potential to unlock capabilities that don't work at smaller scale. This helps developers forecast progress in AI and to tackle more ambitious problems. 

\item
To quantitatively test whether enhancements other than scaling (e.g. hyper-parameter tuning, novel architecture design, etc.) actually matter as models increase in scale. If these non-scale based changes do not give improvements at scale, then developers can allocate developer time to pursuing scale relative to other alternatives.

\item
To debug model training. If a bigger model is not doing better than a smaller model, then developers know to prioritize looking for possible bugs inherent only to models of sufficient scale. Some commonly encountered bugs that become increasingly pernicious with scale involve numerical precision issues, data quality issues, over-fitting issues, and hardware related issues.

\item
To evaluate the performance of models on a common scale. Often, different researchers publish results for models of different sizes. A researcher can use scaling laws to infer how much of the differences in model accuracy are merely due to scale, and also how differently sized models compare to one’s own models after accounting for scale. For instance, an improved approach might be comparable to a $10$\% model size increase. Knowing this information gives two separate options for pursuing such a model improvement.

\end{enumerate}

\subsection{Recommendation System Experiment} \label{app:recsys}

To illustrate how smooth general capability scaling (discussed in Section \ref{sec:1.1}) may correlate with task performance and forecast economic value, we perform a small original experiment where we analyze the relationship between scale and capabilities for GPT-3-like language models \cite{askell_general_2021} to be used as recommendation systems with zero-shot learning. We choose a recommendation system example because these systems have tangible economic relevance and societal impact.  

\begin{figure}
    \centering
    \includegraphics[width=0.33\textwidth]{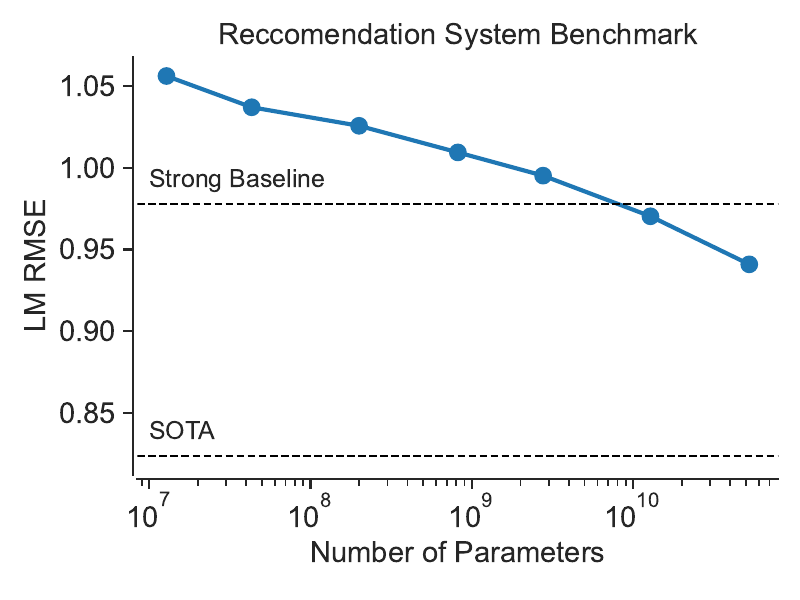}
    \vspace{-1em}
    \caption{Language models can perform as zero-shot recommendation systems with increasing scale. This demonstrates how general capability scaling can correlate with an economically valuable task as described in Section \ref{sec:1.1}.}
    \label{fig:movies}
\vspace{-1em}
\end{figure}

Figure \ref{fig:movies} shows that language models smoothly decrease in the standard Root Mean Square Error (RMSE, lower is better) metric on the widely used Movielens 1M movie recommendation system task \cite{harper_movielens_2015} as they increase in size. The smallest model achieves a significantly better RMSE ($1.06$) than chance (RMSE $1.91$), and the largest model achieves a significantly lower RMSE ($0.94$) than a strong baseline model (RMSE $0.98$, see below for further details). Although no models achieve state of the art (SOTA) performance (RMSE $0.82$), these results are still surprising because the language models (in our zero-shot setting) see two orders of magnitude less training data than the SOTA model.

Trends like those in Figure \ref{fig:movies} forecast how much it would likely cost to develop a state-of-the-art capability on an economically valuable task.  In this particular case, we get an incredulous result - at 800T parameters, a language model will achieve state of the art performance with zero-shot learning. This number indicates that it’s unlikely language models will be used as commercially deployed recommendation systems in this manner for several years (and that even then it might not be worth its costs).\footnote{Of course, algorithmic improvement that shifts the scaling laws is still possible.}  But the results of a different experiment (e.g. a fine-tuned language model trained explicitly to solve this task), could have justified expenditure rather than advising against it. As such, scaling laws can de-risk investment \emph{without saying anything about the detailed behavior of the model in specific} cases. 

More specific technical details are as follows. To perform this experiment, we chose the Movielens 1M (1 million ratings) dataset \cite{harper_movielens_2015} both because of its widespread use, the fact that it contains demographic information about users (age, occupation, gender, zip code), and because we have observed language models to have considerable knowledge about movies (presumably due to a preponderance of text on the internet about movies). 

The dataset consists of $\sim4$K movies rated by $\sim6$K users on a scale of $1$-$5$. On average, each user has rated $\sim~160$ movies, which means $96$\% of the data are missing. The goal of a recommendation system is to predict these missing values, which anticipate how a user will rate a movie they have not previously rated before. 

To evaluate performance on this task, we take the standard approach of partitioning the data into a train and test set, using $1$\% of the total dataset ($10$K ratings) as our test set. Performance on this task is typically reported as the root mean squared error (RMSE) between the predicted and actual ratings on the test set. Perfect predictions would yield an RMSE of $0$ and random guessing corresponds to an RMSE of $1.91$. A strong baseline model simply assigns the average rating (averaged across all users) in the train set as the predicted ratings for all movies in the test set. This essentially ranks movies by their overall popularity, independent of any personalization. The strong baseline achieves an RMSE of $0.98$. State of the art performance on this dataset, is currently an RMSE of $0.822$ according to \cite{han_glocal-k_2021}.\footnote{https://paperswithcode.com/sota/collaborative-filtering-on-movielens-1m}

In general, state of the art algorithms rely on matrix completion (also known as matrix factorization) algorithms, which simply try to impute the missing values in the user-by-movie matrix by expressing this matrix as the outer product of a small number of low dimensional latent vectors, which are learned from the training data, in order to explicitly minimize the RMSE between predicted and actual ratings. Algorithms with lower RMSEs are typically parameterized by neural network models. 

It’s unclear how to use language models as matrix factorizers. Instead, we employ similar zero-shot learning approach with the following prompt:

\begin{spverbatim}
    A {age} {gender} who is employed as an {occupation} previously rated
    {list_of_movies_and_ratings_from_training_set} will rate 
    {movie_from_test_set} a
\end{spverbatim}

\noindent{We replace the variables in curly braces with their corresponding values from the training set (for the first $4$ variables) and test set (for the last variable). We then compute the probability that the language model will predict the next token in the sequence as a  \verb| 1|, \verb| 2|, \verb| 3|, \verb| 4|, or a \verb| 5|. Finally, we compute the weighted mean of the ratings, where the weights are equal to the probabilities (which are normalized to sum to $1$) the language model assigns to those ratings as the final rating prediction. }

We use zero-shot learning because the variable, \verb|list_of_movies_and_ratings_from_training_set|, can often correspond to a very long sequence of text, since on average users have rated $\sim200$ movies. Because our models have a fundamental limit on how large input text sequences can be, as determined by the context window length, we randomly sample up to 500 previously rated movies per user, in order to stay within the limits. An alternative strategy is to use few-shot learning, where multiple users are provided as examples in the prompt, but only show a small number, e.g., $5$, of previously rated movies per user. Empirically, we found that our zero-shot approach (one user, but a lot of previously rated movies by that user) led to far lower RMSE than all of our few-shot learning experiments.

\VerbatimFootnotes
There is a movie recommendation task for language models in the BigBench framework based on a Movielens benchmark.\footnote{https://github.com/google/BIG-bench/blob/main/bigbench/benchmark\_tasks/movie\_recommendation/README.md} However, this task is formulated specifically for a language model, in the form of a multiple choice task as opposed to a regression or matrix completion task. Our formulation of this task allows us to directly compare general purpose language models to special purpose systems on the same axis in order to assess a more realistic capability. Finally, we note that we can extend the analysis we do here to other economically valuable real-world tasks such as those in the recent Real-World Few-Shot Text-Classification (RAFT) benchmark \cite{alex_raft_2021}.

\subsection{COMPAS Experiment} \label{app:compas}

We use the same dataset, filtering operations, and metrics used for the ProPublica analysis of COMPAS \cite{angwin_machine_2016} (taken directly from the Ipython Notebook on GitHub).\footnote{https://github.com/propublica/compas-analysis}
We use the same prompt as  \cite{dressel_accuracy_2018} (which was designed for a human recidivism prediction experiment) for the language model:
\begin{spverbatim}
    The defendant is a {sex} aged {age}. They have been charged with: 
    {c_charge_desc}. This crime is classified as a {c_charge_degree}. 
    They have been convicted of {priors_count} prior crimes. They 
    have {juv_fel_count} juvenile felony charges and {juv_misd_count}
    juvenile misdemeanor charges on their record.
    Do you think this person will commit another crime within 2 years?
\end{spverbatim}

\noindent{We replace the variables in curly braces with values for each defendant in the  dataset. Following \cite{dressel_accuracy_2018} we also consider a prompt in which race is included, by changing the first sentence in the above prompt to be: \newline
\verb|The defendant is a {race} {sex} aged {age}|. We rely on few-shot learning \cite{brown_language_2020}, in which we provide the language model with $50$ labeled examples taken at random from the training dataset (less than $1$\% of the data) before querying the models on each independent test prompt.} 

Next, given the prompt above, we compute the probability that the next token in the prompt is a \verb| Yes| and a \verb| No|. We normalize these two probabilities to sum to 1. We then directly compare the probability of a \verb| Yes|  response to the ground-truth label as to whether or not the defendant in question actually re-offended, in addition to the analogous prediction provided by COMPAS. We use the Fairlearn Python package\footnote{https://github.com/fairlearn/fairlearn} to compute all metrics reported in the main text.

\begin{figure}[t]
    \centering
    \includegraphics[width=0.33\textwidth]{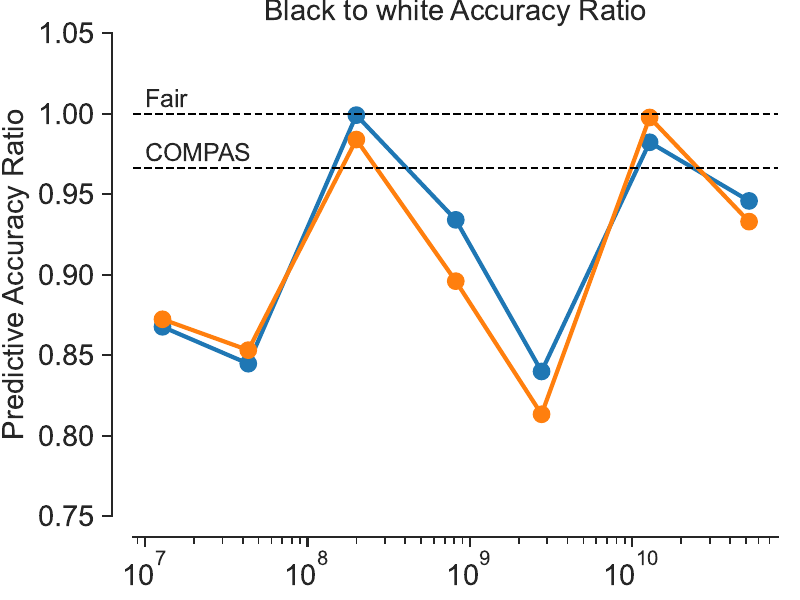}
    \caption{Predictive accuracy ratio for Black versus white defendants. A value of $1$ is fair. COMPAS achieves a value of $0.97$. The language models show no clear trend in this ratio, regardless of when race is excluded in the prompt (blue) or included (orange). However, the largest language models are slightly less equitable than COMPAS according to this metric.}
    \label{fig:predictive_accuracy}
\vspace{-1em}
\end{figure}

In addition to the metrics reported in the main text, we also examined the predictive accuracy ratio for Black versus white defendants as in \cite{angwin_machine_2016, dressel_accuracy_2018}. We saw no clear trends with model size (Figure \ref{fig:predictive_accuracy}) regardless of whether race was excluded from the prompt (blue) or included (orange). Though the largest language models are slightly less fair than COMPAS according to this metric.

Our analysis suffers from several important caveats. First, it is well known that there are many more fairness metrics than the two we consider here, and that it’s statistically impossible for a single algorithm to achieve parity on all these metrics e.g., \cite{friedler_impossibility_2016}. Second, benchmark risk assessment instrument datasets often contain numerous measurement biases and errors which can make them ill-suited for making claims about real-world impact without carefully considering the the complicated socio-technical systems (in this case, the US criminal justice system) in which they are used \cite{bao_its_2021}. Finally, comparisons to proprietary algorithms will always be difficult to make precise without either significant reverse engineering or pressure from companies to lead to more transparent algorithms \cite{rudin_age_2020}. 

\subsection{Open Ended Outputs and Creative Expression}
\label{app:creative_expression}
Capabilities may emerge in areas that are challenging to evaluate quantitatively, and therefore likely to resist systematic analysis.  A key example is the case of AI models mimicking human creative expression.    As a concrete example, we provide\footnote{https://gist.github.com/jareddk/6512393d4a996fbf3a72be265a5285aa} a sample of over three thousand imitation poems generated randomly from a large language model (more accurately, these are samples generated from a prompt including several modern and contemporary poems, so a small fraction of the samples are not actually poems).  We cannot provide any official evaluation, but informally we find both the quality of some of the texts, and the imitation of specific authorial styles quite impressive.  Some professional writers who are aware of the growing capabilities of large language models are very impressed \cite{hoel_big_2021}, but also alarmed by their far-reaching implications. Academics outside of engineering departments are also starting to consider the pros and cons of machine creativity \cite{underwood_science_2021}.

\subsection{Toxicity Experiment Details} \label{app:toxicity}
We follow a similar analysis outlined in \cite{askell_general_2021} where we leverage the RealToxicityPrompts \cite{gehman_realtoxicityprompts_2020} dataset to elicit short comments in response to real world samples of text (prompts) obtained from the internet. Following \cite{gehman_realtoxicityprompts_2020}, we label the prompts as \enquote*{toxic} if they have a toxicity score $>0.5$, otherwise we label them \enquote*{non-toxic}. We then obtain a random sample of $1$K of these prompts, with an equal proportion of \enquote*{toxic} and \enquote*{non-toxic}. Next, we we sample $25$ model responses from language models of various sizes \cite{askell_general_2021} per prompt. We use the same prompts per language model.

We then measure the toxicity of the model responses with an open-source toxicity detector \cite{hanu_detoxify_2020} that outputs a score, between $0$ and $1$, with a higher score corresponding to more toxic content. 
Next, we fit a linear regression model, where we predict the toxicity score based on a categorical coding of model size, and a binary indicator as to whether the prompt was labeled as toxic or non-toxic. We plot the estimated coefficients on model size (thus controlling for the toxicity of the prompt) and the $95$\% confidence intervals around them in the main text.

Our analysis is subject to several caveats. First, it’s unclear how the magnitude of the effect size in Figure \ref{fig:toxicity} influences human perception of the toxicity of the generated text. Different people often have different perceptions about text with the same toxicity score \cite{welbl_challenges_2021}. Second, automated toxicity detection algorithms are known to suffer from several limitations, for example, they can be biased for certain minority groups \cite{gehman_realtoxicityprompts_2020, welbl_challenges_2021}. Finally, our reliance on an open-source toxicity detector \cite{hanu_detoxify_2020} is counter to the more common use of the Perspective API for toxicity detection (though we believe these toxicity detectors to be similar\cite{askell_general_2021}). 

\subsection{AI and Compute Analysis Details} \label{app:ai_compute}
We leverage data from existing work on estimating compute usage for training large-scale AI models\footnote{https://openai.com/blog/ai-and-compute/} which was recently complemented with additional data from more recent experiments \cite{sevilla_parameter_2021}. In this augmented dataset, we label training runs as Industry or Academic based primarily on affiliations of first authors. If a first author had a dual affiliation, we labeled the run as industry, because in practice we’ve found that with access to both, industry-controlled compute is the preferred path. The fit it in Figure \ref{fig:ai_compute} (Right) is based on a LOWESS regression with default parameters from the Seaborn Python package. These data are incomplete and should be interpreted carefully due to sampling bias. For example, we do not have access to compute estimates for industrial models used in production for search, recommendation engines, or self driving cars. 

\end{document}